\documentclass[conference]{IEEEtran}
\IEEEoverridecommandlockouts

\usepackage{cite}
\usepackage{amsmath,amssymb,amsfonts}
\usepackage{algorithmic}
\usepackage{graphicx}
\usepackage{textcomp}
\usepackage{xcolor}

\usepackage{diagbox}
\usepackage{multirow}
\usepackage{multicol}
\usepackage{makecell}
\usepackage{hyperref}

\def\BibTeX{{\rm B\kern-.05em{\sc i\kern-.025em b}\kern-.08em
    T\kern-.1667em\lower.7ex\hbox{E}\kern-.125emX}}
\begin{document}

\title{Automated Detection of Epileptic Spikes and Seizures Incorporating a Novel Spatial Clustering
Prior\\
}

\author{\IEEEauthorblockN{Hanyang Dong$^{1,2,\dag}$, Shurong Sheng$^{2, \dag}$, Xiongfei Wang$^{3, \dag}$, Jia-Hong Gao$^{4}$, Yi Sun$^{2}$,
Wanli Yang$^{2}$, Kuntao Xiao$^{2}$, \\Pengfei Teng$^{3}$, Guoming Luan$^{3,*}$ and Zhao Lv$^{5, *}$ 
}
    \IEEEauthorblockA{$^1$ School of Artificial Intelligence, Anhui University, Hefei 230601, China.\\}
    \IEEEauthorblockA{$^2$ Anhui Province Key Laboratory of Biomedical Imaging and Intelligent Processing, \\Institute of Artificial Intelligence, Hefei Comprehensive National Science Center, Hefei 230088, China.\\}
    \IEEEauthorblockA{$^3$ Department of Neurosurgery, Beijing Key Laboratory of Epilepsy, Sanbo Brain Hospital Capital Medical University, Beijing, China.\\}
    \IEEEauthorblockA{$^4$ McGovern Institute for Brain Research, Peking University, Beijing, China, and Center for MRI Research, Academy for \\ Advanced Interdisciplinary Studies, Peking University, Beijing, China, and Beijing City Key Lab for Medical Physics and Engineering,\\ Institute of Heavy Ion Physics, School of Physics, Peking University, Beijing, China.\\}
    \IEEEauthorblockA{$^{5}$ Anhui Province Key Laboratory of Multimodal Cognitive Computation,\\ School of Computer Science and Technology, Anhui University, Hefei 230601, China.\\} 
\IEEEauthorblockA{$\dag$ Equal contribution, and $*$ Corresponding authors.} 
\IEEEauthorblockA{jgao@pku.edu.cn, shengsr@outlook.com, kjlz@ahu.edu.cn} 
}

%
%

\maketitle

\begin{abstract}

A Magnetoencephalography (MEG) time-series recording consists of multi-channel signals collected by superconducting sensors, with each signal's intensity reflecting magnetic field changes over time at the sensor location. Automating epileptic MEG spike detection significantly reduces manual assessment time and effort, yielding substantial clinical benefits. Existing research addresses MEG spike detection by encoding neural network inputs with signals from all channel within a time segment, followed by classification. However, these methods overlook simultaneous spiking occurred from nearby sensors. We introduce a simple yet effective paradigm that first clusters MEG channels based on their sensor's spatial position. Next, a novel convolutional input module is designed to integrate the spatial clustering and temporal changes of the signals. This module is fed into a custom MEEG-ResNet3D developed by the authors, which learns to extract relevant features and classify the input as a spike clip or not. Our method achieves an F1 score of $94.73\%$ on a large real-world MEG dataset Sanbo-CMR collected from two centers, outperforming state-of-the-art approaches by $1.85\%$. Moreover, it demonstrates efficacy and stability in the Electroencephalographic (EEG) seizure detection task, yielding an improved weighted F1 score of $1.4\%$ compared to current state-of-the-art techniques evaluated on TUSZ, whch is the largest EEG seizure dataset.

\end{abstract}


\section{Introduction}


\begin{figure}
    \centering
    \includegraphics[width=8.8cm]{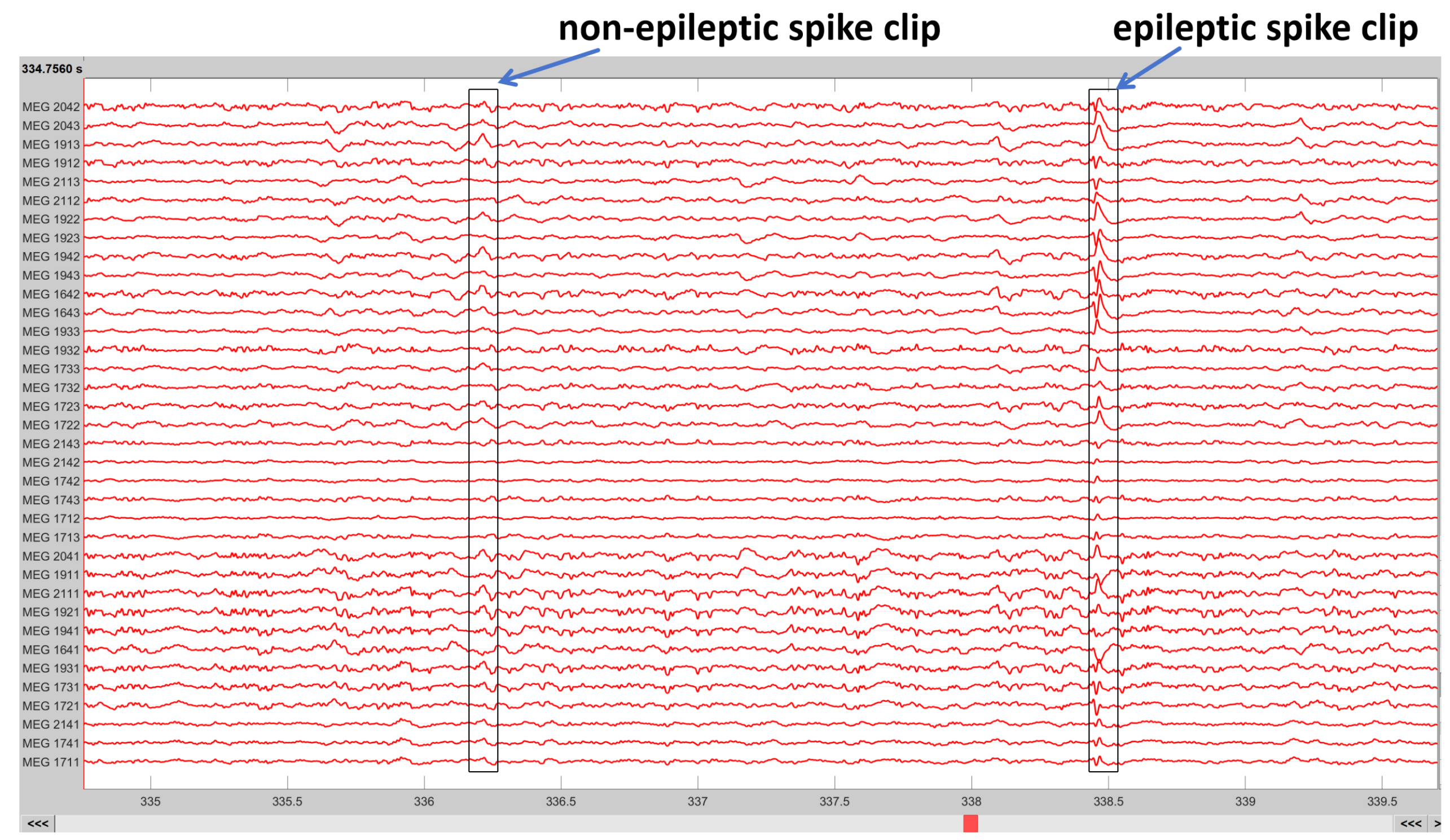}
    \caption{Automatic epileptic MEG spike detection is challenging due to the high similarity between epileptic and non-epileptic spikes, and the regional spiking characteristics.}
    \vspace{-2em}
    \label{fig:spike}
\end{figure}

The primary objectives of epileptic spike and seizure detection are to identify epileptic spikes in MEG data and seizure onset in EEG recordings. These tasks are crucial for seizure analysis, using spikes as biomarkers for epileptogenic zones \cite{moore2002magnetoencephalographically,abd2018review,pan2023magnetoencephalography} and enabling seizure detection to alert professionals during resource-constrained periods \cite{wong2023eeg}. Manual identification of epileptic spikes and seizure onset zones is time-consuming and requires expertise, necessitating the development of an automated, efficient detection algorithm. Although both MEG and EEG recordings are multi-channel signals, an input MEG signal clip has more signal channels than an input EEG clip due to its higher spatial resolution. Moreover, MEG and EEG coincident events exhibit statistically significant differences in several morphological characteristics \cite{fernandes2005does}. Furthermore, the detection of spikes and seizures aims to pinpoint events at the interictal and ictal periods of epilepsy, where their event patterns differ inherently. In this study, we focus on these two tasks to enhance the robustness of our approach across diverse datasets, namely, the Sanbo-CMR MEG spike detection dataset and the Temple University SeiZure detection dataset (TUSZ). Subsequently, we primarily employ epileptic MEG spike detection to introduce the concepts and techniques, with EEG seizure detection provided as an additional application to showcase the spatial-temporal adaptability of the proposed methodology.

Deep learning methods have emerged as the cutting edge in the analysis of seizures in recent years. However, most contemporary approaches perform spike detection by distinguishing spikes from the background based on morphological feature changes over time \cite{hirano2022fully,zheng2019ems,ahmad2023hybrid}. This methodology presents challenges in differentiating epileptic spikes from non-epileptic spikes that arise from noise generated by other human organs yet share similar waveforms \cite{mckay2019artifact}, as illustrated in Fig. \ref{fig:spike}. Moreover, these methods frequently ignore the clustering features that manifest during the interictal phase of absence seizures \cite{richardson2012large,chavez2010functional,ponten2009indications}, which is a crucial aspect in identifying spikes as epileptic in origin. This aspect is termed the “regional spiking characteristic” throughout the subsequent content of this study. In this paper, we introduce a unified framework that addresses these two challenges jointly by designing a novel input module that incorporates the spatial clustering prior for the deep neural network. This module first clusters the multi-channel signals into groups based on their sensor positions and then integrates the 2D clusters with the signal strength variations over time to create a 3D input module. By clustering the signals within a signal clip according to their respective sensor positions, epileptic and non-epileptic spikes are naturally separated into distinct clusters since they originate from different regions. This operation also more effectively encodes the regional spiking characteristics, thereby enhancing spike detection capabilities. Overall, the main contributions of this paper are:\\
(1) A novel input module is designed to reconfigure the input signals within a signal clip according to their spatial clustering priors that can encapsulate spatial clustering patterns, and temporal changes in the signal.\\  
(2) A custom MEEG-ResNet3D is developed to facilitate learning from the novel input module. Additionally, we propose a spatial attention module within this network to encode both intra and inter-cluster attentions, thereby more effectively capturing the spatial clustering patterns of the input. \\
(3) Effectiveness and robustness of our approach are proved through extensive testing on the Sanbo-CMR and TUSZ datasets, which are tailored for detecting spikes and seizures. The TUSZ dataset is the largest public EEG seizure dataset, covering a variety of seizure types and a broad patient population. Thorough experiments are conducted to analyze our approach for both spike and seizure detection, providing insights for future research.

\section{Related Work}
\label{sec:related_work}

\subsection{{Epileptic MEG/EEG Spike Detection}}
The vast majority of deep learning approaches for spike and seizure detection involve distinguishing these events from the background by analyzing patterns in the frequency or time domain. For instance, Zheng et al. \cite{zheng2019ems} begin by extracting features related to local and global signal variations using 1D and 2D convolutional neural networks (CNNs) for single and multi-channel signals, respectively, and then apply these features for spike classification. The essence of their method lies in differentiating spikes from the background by examining the morphological changes of the signals over time. WSNet \cite{ABDULWAHHAB2024114700} focuses on seizure detection through the extraction of frequency-domain features using a long short-term memory neural network. The core idea behind WSNet is to discern the frequencies specific to seizures from the background. Some approaches concurrently investigate spike or seizure detection by encoding signal variations in both temporal and spatial dimensions. For example, Ming et al. \cite{MING2024106136} employ a dual-tower network to extract features from both the time and spatial domains of the signals; however, the strict adherence of the input spatial matrix to the layout of EEG sensors limits its adaptability to other tasks involving MEG. Additionally, research on spike or seizure detection incorporating multi-level features exists. Cheng et al. \cite{cheng2022multilevel} perform EEG spike detection utilizing a variety of features, including those from the time domain, spatial domain, and frequency domain. Nevertheless, this approach faces difficulties in modeling the overlapping of intra-domain and inter-domain features and formulating a clear and comprehensive multi-domain representation for classification purposes. In contrast, our study introduces a simple yet effective model that can capture the spatial-temporal and clustering information of input signals using a unified 3D convolutional neural network.
\subsection{Spatial Relation Encoding for Multivariate Time-series}
Research in the encoding of spatial relationships for Multi-Channel Time-Series (MTS) data has primarily focused on architectures based on 3D CNNs or Transformers. The 3D CNN-based approach aims to formulate an EEG representation that effectively preserves both temporal and spatial information. For instance, 3DCANN \cite{Liu_Wang_Zhao_Li_Hu_Yu_Zhang_2022} employs 3D CNN blocks to capture spatial-temporal features, which are then combined with a dual attention module to enhance performance in EEG emotion recognition tasks. Due to their proficiency in encoding multivariate relationships, Transformers are also utilized for spatial relation encoding. Crossformer \cite{zhang2022crossformer} exemplifies this by attempting to encode cross-temporal and cross-dimensional information for MTS forecasting. Specifically, the input MTS is converted into a 2D vector array through a Dimension-Segment-Wise embedding module to maintain time and dimension information. This is followed by the introduction of a Two-Stage Attention layer to efficiently capture dependencies across time and dimensions for the final MTS forecast. Crossformer can be adapted to address the MEG spike detection problem by encoding channel-wise and time-wise features in the input multi-channel signals, utilizing channel-wise positional embedding to capture relative channel positions. Additionally, custom-designed 3D-CNN and Transformer based models have been used in other research (e.g., \cite{Sun_Wang_Zhao_Hao_Wang_2022,liu2023itransformer,Liu_Yang_2021}) to encode spatial relationships. Distinct from the aforementioned methods, DCRNN-GNN \cite{tang2021self} employs a graphical neural network to encode the spatial topography of EEG sensors, followed by the use of a Long Short-Term Network (LSTM) to extract spatial-temporal features from the input signals. We will compare our model architecture with these three types of architectures to validate the effectiveness of the neural network we propose.

\section{Methodology}
\label{sec:methodology}
\begin{figure*}[h!]
    \centering
    \includegraphics[width=17cm]{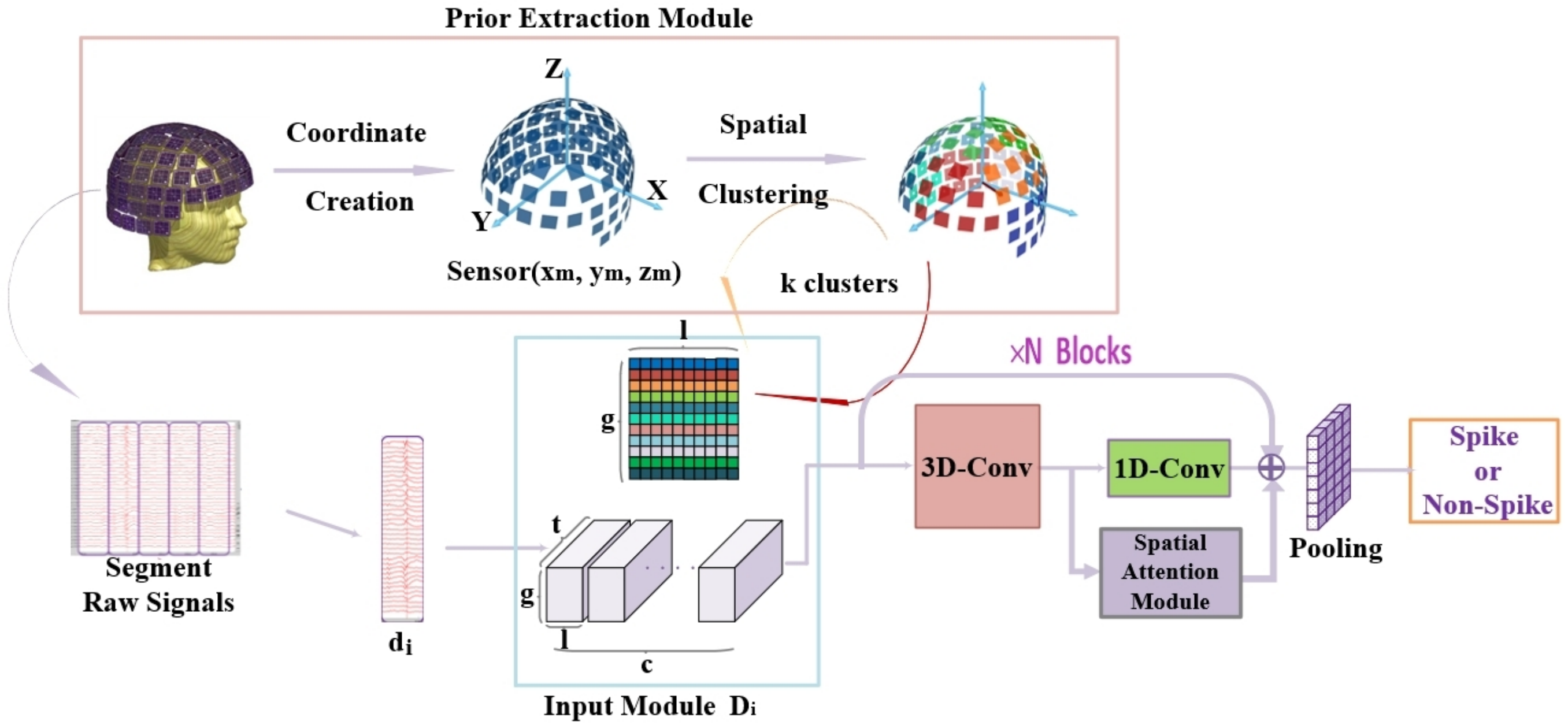}
    \caption{Overview of the paradigm. The spatial and clustering priors are integrated into a self-designed 3D input module, which also incorporates signal changes within a time clip. Subsequently, a custom  MEEG-ResNet3D is developed to extract relevant knowledge from the input and perform spike or seizure classification.}
    \label{fig:workflow}
\end{figure*}
In this section, we denote matrices by using bold font to enhance the clarity of the symbols within this study. For each MEG recording $\boldsymbol{d}$ in the dataset, the objective is to identify whether a signal clip $\boldsymbol{d_i}$ segmented from $\boldsymbol{d}$ contains an epileptic spike or not. The overview of our classification paradigm is illustrated in Fig. \ref{fig:workflow}. This paradigm consists of three steps. In the first step, prior information and signal segments are extracted and processed. Next, an input module is designed to encode the priors and signal variations over time. In the third step, this input module is fed into a MEEG-ResNet3D network for classification. We will provide detailed explanations of these steps in the following sections.

\subsection{Prior Extraction and Input Signal Processing}
\label{prior}
\subsubsection{Spatial and Clustering Prior Extraction}
The abstract highlights that for a MEG recording $\boldsymbol{d}$, we are dealing with multi-channel signals generated by sensors placed on the head. We can think of the sensor locations as a set $S$ of points $s_m$, where each point $s_m$ has coordinates $x_m$, $y_m$, and $z_m$ in Euclidean space. To organize these points, we use the k-means clustering algorithm \cite{likas2003global} to group $S$ into $g$ clusters in this step, resulting in $S = \{S_1, ..., S_g\}$.
\subsubsection{Input Signal Processing}
We adopt the preprocessing methods introduced in EMSNet \cite{zheng2019ems} to process MEG data and the techniques presented in DCRNN-GNN \cite{tang2021self} for EEG data. This involves several steps: first, we sample all data across a dataset to a unified sampling rate. Next, we apply filters such as band-pass (3-80Hz) and notch filters (50Hz) to eliminate industrial and environmental noise from the signals. Then, we apply the FastICA algorithm to remove the disturbance obtained from ElectroCardioGraphy (ECG) and Electro-OculoGram (EOG) artifacts. Finally, we segment the cleaned signals into clips of a fixed duration. After preprocessing, an MEG recording $\boldsymbol{d}$ with $T$ time points is divided into $T/t$ clips, each clip $\boldsymbol{d_i}$ being a segment with $t$ sampling points. The duration $t$ is selected considering the distribution of spike or seizure durations.

\subsection{Input Encoding}
\label{input encoding}

The preprocessing phase yields the input signal clip $\boldsymbol{d_i}$ $\in$ ${\mathbb{R}^{n\times t}}$ along with $g$ spatially proximate sensor clusters. Here $i$ denotes the index of this sample within the dataset and $n$ represents the number of channels in $\boldsymbol{d_i}$. It is evident that $\boldsymbol{d_i}$ only captures the temporal signal variations from multiple channels. Since the channels in $\boldsymbol{d_i}$ are produced by sensors, we can accordingly partition the signals in $\boldsymbol{d_i}$ into $g$ clusters based on the sensors from which they originate. To integrate the spatial clustering information extracted from Section \ref{prior}, we reorganize the first dimension of $\boldsymbol{d_i}$ into a 2D spatial-cluster map based on the affiliation between each channel and its corresponding cluster. More specifically, we define a 3D input $\boldsymbol{{D_i}} = [\boldsymbol{h^1_i},...,\boldsymbol{h^j_i},...,\boldsymbol{h^t_i}]$, where $\boldsymbol{h^j_i}$ with a dimension of $\mathbb{R}^{g\times l}$ represents the spatial-cluster map. Here $g$ is the number of clusters mentioned earlier and $l$ represents the maximum amount of sensors within a cluster. The values in $\boldsymbol{h^j_i}$ indicate the signal strength at the sampling point $j$. If a single sensor at a specific location generates multiple channel signals, we can formulate $\boldsymbol{D_i}$ into a shape of $(g,l,t,c)$ where $c$ represents the number of signals produced by each sensor. The process to convert the 2D input $\boldsymbol{d_i}$ to the target 3D input $\boldsymbol{D_i}$ is illustrated in Fig. \ref{fig:workflow}.

\begin{figure}[tb]
    \centering
    \includegraphics[width=7.5cm]{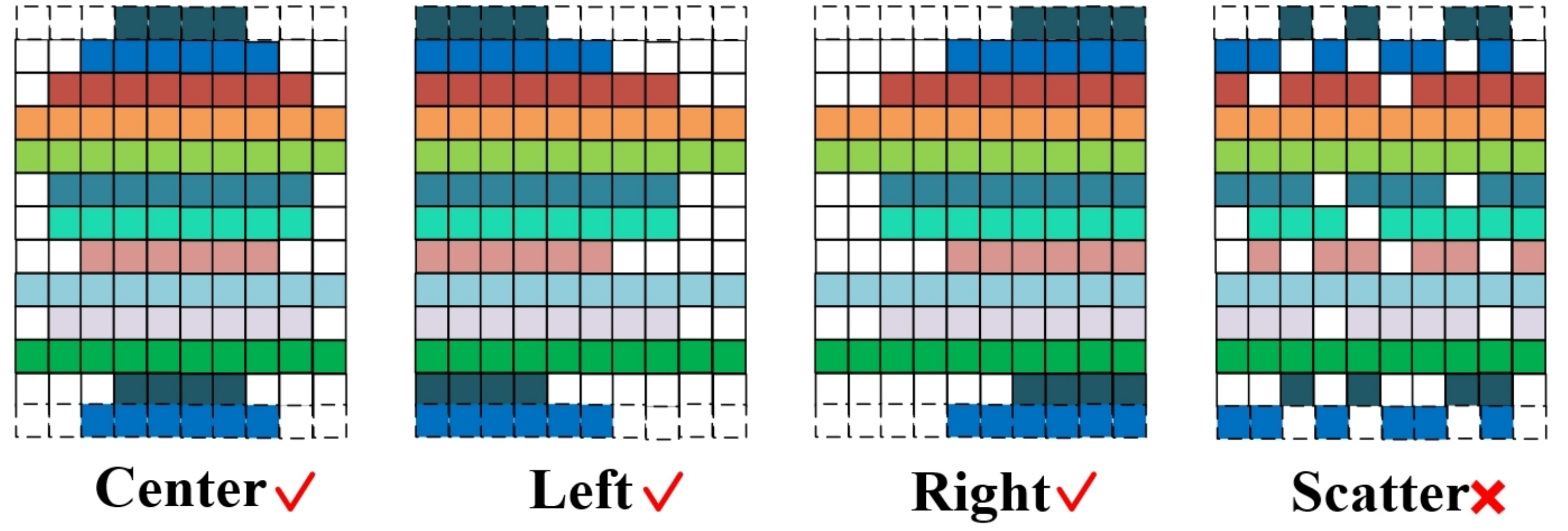}
    \caption{Illustration of the padded spatial arrangements for the spatial-clustering map in the input. }
    \vspace{-1.5em}
    \label{fig:diff-permute}
\end{figure}

The input $\boldsymbol{D_i}$ is padded with the circular padding approach at the top and bottom boundaries to allow the model to connect sensors that are far apart. Conversely, zero padding is used to fill in the clusters with fewer sensors than $l$. This input method gives us substantial flexibility in how we arrange sensors or channels within a cluster in the spatial clustering map. As shown in Fig. \ref{fig:diff-permute}, we choose to place sensors or channels densely along the left, right, or center of a cluster, rather than spreading them out. This dense arrangement is more effective for convolutional kernels to extract features compared to a scattered approach. We also try different permutations between clusters to make our study more thorough.  

\subsection{Network Architecture}

\subsubsection{Overall Structure}
Given the lack of existing 3D CNN architectures specifically tailored for MEG spike detection, we have designed a 17-layer MEEG-ResNet3D architecture, which is inspired by 3D ResNet, to address this challenge. Details of this architecture are presented in Table \ref{tab:resnet}. In this table, the kernel with a size of $1\times 1\times n$ indicates that the kernel applies 1D-convolution 
along the time dimension with a fixed size $n$. The kernel configuration within each block allows for a comprehensive analysis of both spatial clustering and temporal aspects of the input. The digits within the stride tuple correspond to the dimensions parallel to those in the kernel size. The output size does not include the batch size $b$ in Table \ref{tab:resnet} as it remains constant throughout the model’s computation. Additionally, to bolster the network’s ability to capture comprehensive information and reduce information loss at deeper layers, the number of depth maps incrementally grows with the network’s depth.  Different signals generated by a sensor share the same kernel weights. Furthermore, we have integrated a spatial attention module into our MEEG-ResNet3D to more effectively capture spatial-cluster information. Details of this spatial attention module are provided subsequently.

\subsubsection{Spatial attention module}
As depicted in Fig. \ref{fig:workflow}, the spatial attention module is applied to the feature maps obtained from the $3 \times 3 \times 3$ convolutional kernel within each block of the network. Specifically, for each feature map $\boldsymbol{F^q_{i}}$ at layer $q$, which is of dimension ${\mathbb{R}^{g^q \times l^q \times t^q \times c^q}}$, we compute the attention matrix in three steps. First, we average $\boldsymbol{F^q_{i}}$ across the temporal and feature-depth dimension, resulting in an integrated spatial-cluster map $\boldsymbol{f^q_i}$ of dimension ${\mathbb{R}^{g^q \times l^q }}$ that aggregates information from all sampling points across all depths. Next we apply a softmax operation along the second dimension of $\boldsymbol{f^q_i}$ to compute the attention scores for each channel within a cluster. These initial scores are sorted in descending order and then scaled with the maximum value of respective cluster in $\boldsymbol{F^q_{i}}$. This step produces a transformed spatial-cluster map $\boldsymbol{f^q_i}'$ $\in$ ${\mathbb{R}^{g^q \times l^q }}$ that captures intra-cluster attention. Finally, we apply a softmax operation to the first dimension of $\boldsymbol{f^q_i}'$ to compute the inter-cluster attention scores for the spatial-cluster map. Through these three steps, we generate a weighted matrix $\boldsymbol{w^q_i}$ that embodies both intra-cluster and inter-cluster attentions. We then rearrange the elements of $\boldsymbol{w^q_i}$  along the second dimension to align with the order of $\boldsymbol{F^q_{i}}$ for subsequent operations, specifically, element-wise multiplication with the feature map derived from the $1 \times 1 \times n$ kernel in layer $q$. The computation of $\boldsymbol{w^q_i}$ can be represented by the following equations:

\vspace{-0.4cm}
\begin{align}
    \boldsymbol{f^q_i} &= \frac{1}{t^q\times c^q}\sum_{c,t}\boldsymbol{F^q_i}
\end{align}
\begin{align}
    \boldsymbol{{f^q_i}'} &= [\frac{e^{m_{g',l'}}}{\sum_{g',*} e^{m_{g',*}}}]\odot \mathop{\max_{l'}}
 \boldsymbol{F^q_{i}} & g'\in[1,g^q], l'\in[1,l^q]
\end{align}

\begin{align}
    \boldsymbol{{f^q_i}'} &= \mathop{Descending Sort} \boldsymbol{{f^q_i}'}\\
    \boldsymbol{w^q_i} &= [\frac{e^{m'_{g',l'}}}{\sum_{*,l'} e^{m'_{*,l^q}}}] & g'\in[1,g^q], l'\in[1,l^q]
\end{align}

Where $m_{g',l'}$ and $m'_{g',l'}$ represent the element values of $\boldsymbol{f^q_i}$ and $\boldsymbol{{f^q_i}'}$ respectively. The vector obtained with the $\mathop{\max_{l'}}
 \boldsymbol{F^q_{i}}$ operation contains the maximum value of each cluster from the spatial clustering map within $\boldsymbol{F^q_{i}}$.
 
\subsection{Learning Objective}
In this paper, we approach the spike detection problem as a binary classification task. Specifically, given an input signal clip $d_i$, suppose its ground-truth label is $y_i$, which takes a value of 0 or 1, the objective can be formulated as follows:
\begin{equation}
\begin{split}
    l =-\frac{1}{N}\sum_{i=1}^{N}\big(y_{i} log(p_{i}) + (1 - y_{i})log(1-p_{i})\big) 
\end{split}
\end{equation}
Where $p_i$ denotes the prediction value made by the model, and $l$ refers to the binary cross-entropy loss. $N$ represents the number of training samples within a mini-batch. 

\begin{table}[tbp]  
    \centering  
    \caption{Network architecture of the 17-layer MEEG-ResNet3D model in MEG spike detection.} 
     \resizebox{9cm}{2.8cm}{
    \begin{tabular}{ccc}
    \hline
         \textbf{Layer name} & \textbf{Block type} & \textbf{Output size} \\ \hline
         Input size& \multicolumn{2}{c}{13$\times$10$\times$300$\times$3} \\ \hline
         Conv1 & 1$\times$1$\times$5, 64, stride (1,1,2) & 13$\times$10$\times$150$\times$64 \\ \hline
         Res-block1 & \makecell*{$\begin{bmatrix} 3 \times3\times3, 64, \rm stride(1,1,2) \\
                        1\times1\times3, 64, \rm stride(1,1,1)   \end{bmatrix}$} $\times$2 & 13$\times$10$\times$75$\times$64\\ \hline
         Res-block2 & \makecell*{$\begin{bmatrix} 3 \times3\times3, 128,\rm stride(1,1,2) \\
                        1\times1\times3, 128, \rm stride(1,1,1)   \end{bmatrix}$} $\times$2 & 13$\times$10$\times$38$\times$128\\ \hline
        Res-block3 & \makecell*{$\begin{bmatrix} 3 \times3\times3, 256, \rm stride(1,1,2) \\
                        1\times1\times3, 256,\rm stride(1,1,1)   \end{bmatrix}$} $\times$2 & 13$\times$10$\times$19$\times$256\\ \hline
        Res-block3 & \makecell*{$\begin{bmatrix} 3 \times3\times3, 512,\rm stride(1,1,2) \\
                        1\times1\times3, 512, \rm stride(1,1,1)   \end{bmatrix}$} $\times$2 & 13$\times$10$\times$10$\times$512\\ \hline
        Average-pool & 13$\times$10$\times$10 & 1$\times$1$\times$1$\times$512 \\ \hline
        FC \& Softmax & 512$\times$2 & 2 \\  \hline
        FLOPs \& Trainable params  & \multicolumn{2}{c}{FLOPs: $13.78 \times 10^9$,  params: $16.26 \times 10^6$ } \\ \hline
    \end{tabular}
    }
    \label{tab:resnet}
\end{table}

\section{Experiments}
\subsection{Datasets}
For the task of \textbf{detecting epileptic MEG spikes}, we assess the performance of different methods using a private dataset \textit{Sanbo-CMR}. This dataset consists of 306-channel MEG recordings collected from the Sanbo Brain Hospital of Capital Medical University and the CMR Center of Peking University. Our study is approved by the Institutional Review
Board of both centers mentioned above and written informed consent was obtained from all patients. The \textit{Sanbo-CMR} dataset extends the \textit{EMS} dataset used in EMS-Net \cite{zheng2019ems}, incorporating additional samples from Sanbo Hospital and also including samples from the CMR Center of Peking University. The spike events in the MEG recordings within \textit{Sanbo-CMR} have been meticulously annotated by a leading neurophysiologist and verified by two additional neurophysiologists. Here, a single spike event is considered to be existence of any spike across all channels within a brief time frame and it is important to note that MEG recordings are divided into multiple non-overlapping segments, each lasting 300 ms. The training and test sets are selected from different patients to assess the generalization ability of each model, while maintaining a nearly balanced ratio of positive to negative labels and incorporating challenging hard cases that are difficult to identify.

For the \textbf{EEG seizure detection} task, we utilize the TUSZ dataset \cite{shah2018temple} v2.0.0, which is the largest public EEG dataset available for seizure detection (\url{https://isip.piconepress.com/projects/tuh_eeg/}). In this study, we incorporate 19 EEG channels and exclude five patients who are included in both the official TUSZ training and testing sets, in alignment with the state-of-the-art (SOTA) method. 
A clip is labeled as “seizure” if it contains any seizure-like segments. The training set is nearly balanced, with data marked as seizures constituting 6.98\% and 9.53\% of the validation and test sets, respectively. Table \ref{tab:dataset_illustrate} illustrates the statistical information of two datasets.

\begin{table}[b]
    \centering
    \caption{Statistics of the two datasets utilized in this study}
    \resizebox{9.0cm}{0.5cm}{
    \begin{tabular}{cccccc}
    \hline
        \textbf{Datasets} & \textbf{Subjects (train/val/ test)} & \textbf{Samples} & \textbf{Channels} & \textbf{Clip length} & \textbf{Classes} \\ \hline
         Sanbo-CMR & 106 / 13 / 13 & 26,713 / 2,374 / 2,332 & 306 & 300 ms & 2 \\ \hline
         TUSZ & 579 / 53 / 43 & 32,706 / 68,864 / 36,497 & 19 & 2400 ms & 2 \\ \hline
    \end{tabular}
    }
    \label{tab:dataset_illustrate}
\end{table}

\subsection{State-of-the-art Approaches}
In the \textbf{MEG spike detection} task, to evaluate the effect of proposed network, our model is compared with seven contemporary and SOTA  methods. These include: (1) two 1D and 2D CNN-based models, EMSNet \cite{zheng2019ems} and SpikeNet \cite{jing2020development}, which detect spikes by differentiating their morphological features from non-spikes using CNNs; (2) Satelight \cite{fukumori2022satelight}, a spike detection model which combines a 2D CNN with a self-attention mechanism; (3) ResNet3D-18 \cite{Hara_Kataoka_Satoh_2017} and TB3D \cite{Liu_Yang_2021} are two 3D CNN networks akin to our MEEG-ResNet3D, 
 which we have incorporated the spatial clustering prior proposed in our study; (4) two transformer-based models adapted for MEG spike detection, Crossformer \cite{liu2023itransformer} and iTransformer \cite{liu2023itransformer}, capable of encoding channel-wise and temporal relationships, utilizing position encodings to capture the relative positions between different channels, enabling the learning of feature clusters during training process. Their approach shares a similar concept with the underlying principle of our methodology.

For the \textbf{EEG seizure detection} task, our method is benchmarked against six contemporary SOTA methods. These include: (1) WSNet \cite{ABDULWAHHAB2024114700}, an advanced seizure detection model that differentiates between seizure and non-seizure states based on frequency features; (2) two 3D CNN-based EEG emotion recognition models repurposed for seizure detection: 3DCANN \cite{Liu_Wang_Zhao_Li_Hu_Yu_Zhang_2022}, and TB3D \cite{Liu_Yang_2021}, which encode the spatial positions of EEG sensors corresponding to their head layout; (3) MMM \cite{yi2023learning}, a transformer-based model tailored for seizure detection through time series forecasting. It effectively maps the Euclidean coordinates of sensor values to positional encodings. (4) V2IED \cite{MING2024106136}, a dual-tower CNN model designed for EEG spike detection, customized for seizure identification. This model not only encodes the spatial positions of sensors corresponding to their head arrangement but also utilizes 2D CNNs to extract both morphological and spatial features pertinent to EEG spike detection. (5) DCRNN-GNN \cite{tang2021self}, an innovative seizure detection model that leverages a graphical neural network to encode the spatial and temporal relations of EEG sensors. It then applies the LSTM to extract spatial-temporal features from the input signals for seizure analysis. The above baselines are carefully selected to evaluate the effectiveness of our network architecture, with the exception of WSNet.

\subsection{Implementation Details}

For the MEG spike detection models, we implement our framework in PyTorch \cite{ketkar2021introduction}, utilizing a mini-batch size of 64. During the backpropagation process, we employ the AdamW optimizer \cite{Loshchilov2017DecoupledWD}. The initial learning rate is semantically set to 0.005 and is progressively decayed by epochs using a cosine-annealing strategy. The number of spatial clusters $g$, as introduced in Sections \ref{prior} and \ref{input encoding}, is determined through cross-validation to be 13, we refer the readers to section \ref{sec:ablation-study} for a detailed illustration of the cross validation results. Additionally, we fix $l$ at 10, which corresponds to the maximum number of sensors within each cluster. For the EEG seizure detection models, we maintain the same hyperparameters as those used for the MEG spike detection task, except for $g$, which is set to 4, and $l$, which is set to 5. 


\subsection{Baseline Implementation Details}


For CNN-based baseline models, such as the 2D-CNN baselines EMSNet \cite{zheng2019ems} and Spike-Net \cite{jing2020development}, we modified the input channels to 306 to align with our model’s input requirement and expanded original kernel sizes to enhance their performance on our task. 
Regarding the 3D-CNN models \cite{Liu_Yang_2021, Hara_Kataoka_Satoh_2017, Liu_Wang_Zhao_Li_Hu_Yu_Zhang_2022, MING2024106136}, we maintained their original architectures as they were already compatible with our 3D input modules. For transformer-based models originally designed for prediction tasks \cite{zhang2022crossformer, liu2023itransformer}, we removed the decoders to adapt them for our spike detection task. For \cite{fukumori2022satelight} and \cite{yi2023learning} which are based on the transformer architecture and originally designed for spike and seizure detection tasks, we preserved their original hyperparameter settings. For the baseline DCRNN-GNN \cite{tang2021self}, we replicated the same model architecture and hyperparameters as reported in the original study. This is because the baseline is specifically designed for seizure detection, and has already been evaluated on the TUSZ dataset.

\subsection{Evaluation Metrics}
We evaluate the performance of the spike and seizure detection models through the assessment of Sensitivity (TPR), Specificity (TNR), F1 Score, and Area Under the Receiver Operating Characteristic Curve (AUROC) metrics. TPR and TNR are crucial for medical models as they reflect the model’s accuracy in identifying true positives and true negatives, respectively. When dealing with imbalanced test sets, where the positive and negative classes are not represented equally, the F1 Score and AUROC provide a comprehensive evaluation of the model’s performance by considering both precision and recall, as well as the model’s overall classification capabilities.  

\section{Results and Analysis}
\label{result-analysis-meg}
\begin{table}
    \small
	\centering 
	\caption{Performance comparison of different models for MEG spike detection evaluated on Sanbo-CMR dataset with four metrics.}
  \resizebox{9.0cm}{1.4cm}{
	\begin{tabular}{lcccc}
		\hline 
            \textbf{Model} & \textbf{TPR (\%)} & \textbf{TNR (\%)} & \textbf{F1 (\%)} & \textbf{AUROC} \\ \hline
            EMSNet\cite{zheng2019ems} & 91.28 & 91.22 & 91.24 & 0.972  \\\hline
            SpikeNet\cite{jing2020development} & 87.24 & 87.28 & 87.22 & 0.927 \\ \hline
            Satelight\cite{fukumori2022satelight} & 76.77 & 76.53 & 75.86 & 0.917 \\ \hline
            ResNet3D18\cite{Hara_Kataoka_Satoh_2017}(w/ clus prior) & 92.88 & 93.94 & 92.88 & 0.929  \\ \hline
            TB3D\cite{Liu_Yang_2021}(w/ clus prior) & 80.68 & 93.34 & 80.41 & 0.865 \\ \hline
            iTransformer\cite{liu2023itransformer} & 79.80 & 79.92 & 79.28 & 0.924  \\ \hline
            Crossformer\cite{zhang2022crossformer} & 81.41 & 81.47 & 81.72 & 0.888 \\ \hline
            Ours & \textbf{94.73} & \textbf{94.73} & \textbf{94.73} & \textbf{0.988} \\ \hline
	\end{tabular}
 }
	\label{spike_detection_results}
\end{table}

\begin{table}
    \small
	\centering 
	\caption{Performance comparison of different models for EEG seizure detection evaluated on TUSZ dataset with four metrics} .
  \resizebox{9.0cm}{1.3cm}{
	\begin{tabular}{lcccccc}
		\hline 
            \textbf{Model} & & & \textbf{TPR (\%)} & \textbf{TNR (\%)} & \textbf{F1 (\%)} & \textbf{AUROC} \\ \hline
            WSNet\cite{ABDULWAHHAB2024114700} & & & 87.8 & 87.7 & 89.3 & 0.834  \\\hline
            TB3D\cite{Liu_Yang_2021} & & & 85.4 & 89.7 & 87.7 & 0.793 \\ \hline
            CANN3D\cite{Liu_Wang_Zhao_Li_Hu_Yu_Zhang_2022} & & & 78.9 & 87.2 & 83.1 & 0.749 \\ \hline
            MMM\cite{yi2023learning} & & & 80.2 & 88.6 & 84.4 & 0.835 \\ \hline
            V2IED\cite{MING2024106136} & & & 88.2 & 88.1 & 87.1 & 0.500  \\ \hline
            DCRNN-GNN\cite{tang2021self} & & & 88.3 & 90.8 & 89.4 & 0.826 \\ \hline
            Ours & & & \textbf{93.0} & \textbf{93.0} & \textbf{93.2} & \textbf{0.894} \\ \hline
	\end{tabular}
 }
	\label{seizure_detection_results}
\end{table}

Table \ref{spike_detection_results} and table \ref{seizure_detection_results} present the outcomes of various models developed in this study for MEG spike detection and  EEG seizure detection tasks. It is evident that our MEEG-ResNet3D model demonstrates superior performance and significantly outperforms all baseline methods across all evaluation metrics on both tasks.
In the context of MEG spike detection task, our method exceeds the performance of the leading baseline method, ResNet3D-18 which incorporates the spatial clustering prior as our model, by 1.85\%  for the \textit{Sanbo-CMR} MEG dataset; For the EEG seizure detection task, our method surpasses the performance of the leading baseline method, DCRNN-GNN, by 3.80\% for the \textit{TUSZ} EEG dataset, both evaluated on F1 score. This clearly demonstrates the advantages of the proposed method in detecting epileptic spikes and seizures due to its ability to capture both spatial clustering and temporal changes of the input signals. Furthermore, 
the effectiveness of our network architecture is evidenced by its superior performance compared to other 3D convolutional neural networks, such as ResNet3D-18 and TB3D, when enhanced with our spatial clustering prior. 

\begin{figure}[b]
    \centering
    \includegraphics[width=\columnwidth]{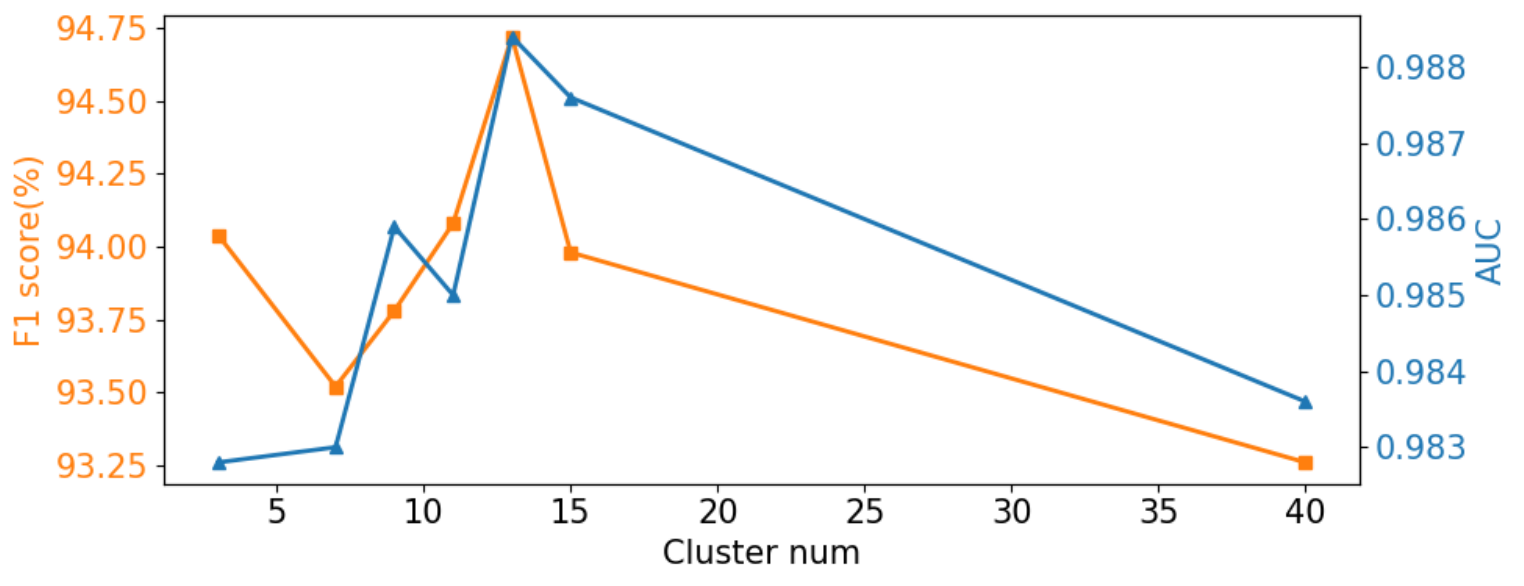}
    \caption{Performance variation of our model on MEG spike detection along the number of clusters: assessment via F1 Score and AUROC.}
    \label{fig:K-cross-val}
\end{figure}

\subsection{Ablation Studies}
\label{sec:ablation-study}
In this section, we conduct ablation studies to demonstrate the effectiveness of the spatial clustering prior introduced in this study, as well as to validate the value chosen for the critical parameters of our model.

To validate the efficacy of the spatial clustering prior, we compare our model with two alternative approaches. In the first approach, we modify the spatial clustering prior by randomly permuting the channels to eliminate the prior information. We then average the F1 scores from five iterations of these random permutations to account for potential discrepancies and improve the comparability of the results. In the second approach, we contrast our model against a previous method where a single EEG/MEG signal input is organized into eight groups, each containing approximately the same number of channels, aligning with the human brain’s functional divisions. The outcomes of these experiments are compiled in Table \ref{prior contrast results}. Across both the balanced \textit{Sanbo-CMR} MEG spike detection dataset and the imbalanced seizure detection dataset \textit{TUSZ}, the MEEG-ResNet3D architecture that integrates spatial clustering priors exhibits a marked enhancement in performance compared to its counterpart devoid of such priors, as well as those utilizing functional brain division priors.

The hyperparameters that significantly influence model performance include the number of clusters $g$, the depth of our MEEG-ResNet3D architecture, and the permutation strategies utilized within and between clusters. Fig. \ref{fig:K-cross-val} depicts the performance fluctuations across varying values of $g$. The graph clearly shows that the model’s effectiveness significantly declines when $g$ deviates from the optimal value of 13. Therefore, 13 is uniformly adopted as the grouping parameter in all experiments designed for the MEG spike detection task. Table \ref{tab:hyperparams} outlines the model’s performance with different network depths, revealing that the 17-layer network outperforms both its deeper and shallower counterparts. When exploring intra-cluster permutations of input channels, the spike detection task on the \textit{Sanbo-CMR} dataset achieves peak performance of 94.73\% and a lowest of 93.91\%. For the seizure detection task on the \textit{TUSZ} dataset, the corresponding figures are 93.2\% and 93.0\%. Notably, the seizure detection task is limited to left and right permutations due to the substantial reduction in the number of channels in the EEG input. Regarding inter-cluster permutations, the average F1 scores for the spike and seizure detection tasks are 94.35\% and 91.7\%, respectively. Thus, even when altering both intra-cluster and inter-cluster permutations, the lowest and average performances of our model remain superior to those of the SOTA baseline approaches.

\begin{table}[b]
    \small
	\centering 
	\caption{Model performance with and without spatial-clustering priors, and with brain functional division priors.}
  \resizebox{9.0cm}{1.1cm}{
	\begin{tabular}{lccccc}
		\hline 
		\textbf{Datasets} & \textbf{Model Prior}
		&\textbf{TPR (\%)} &\textbf{TNR (\%)}&\textbf{F1 (\%)}&\textbf{AUROC}   \\ \hline
		 \multirow{3}*{Sanbo-CMR} & -- & 93.60$\pm$0.290 & 92.87$\pm$0.569 & 93.60$\pm$0.287 & 0.985$\pm$0.0005  \\
         & functional sector & 94.00 & 93.00 & 94.00 & 0.983  \\
		 & clus priors & \textbf{94.73} & \textbf{94.73} & \textbf{94.73} & \textbf{0.988}  \\ \hline
            \multirow{3}*{TUSZ} & -- & 87.7$\pm$2.39 & 90.92$\pm$1.05 & 89.6$\pm$1.68 & 0.871$\pm$0.02  \\
             & functional sector & 90.1 & 92.0 & 91.1 & 0.881  \\
             & clus priors& \textbf{93.0} & \textbf{93.0} & \textbf{93.2} & \textbf{0.894}  \\ \hline
	\end{tabular} 
}
	\label{prior contrast results}
\end{table} 

 \begin{table}[b]
    \small
	\centering 
	\caption{Performance of the MEEG-ResNet3D with different hyper-parameters.}
  \resizebox{9.0cm}{1.5cm}{
	\begin{tabular}{lccccc}
		\hline 
		\textbf{Datasets} & \textbf{Model}
		&\textbf{TPR ($\%$)} &\textbf{TNR ($\%$)}&\textbf{F1 ($\%$)}&\textbf{AUROC}   \\ \hline 
		
		\multirow{3}*{Sanbo-CMR} & MEEG-ResNet3D-9 & 93.70 & 93.34 & 93.70 & 0.981   \\
                            ~ & MEEG-ResNet3D-13 & 94.09 & 93.00 & 94.08 & 0.983 \\
                            ~ & MEEG-ResNet3D-17 &\textbf{94.73} & \textbf{94.73} & \textbf{94.73} & \textbf{0.988}  \\
                            ~ & MEEG-ResNet3D-21 & 94.42 & 94.62 & 94.43 & 0.985 \\ \hline
        \multirow{3}*{TUSZ} & MEEG-ResNet3D-9 & 86.9 & 90.5 & 89.1 & 0.895  \\
                            ~ & MEEG-ResNet3D-13 & 91.2 & 92.7 & 92.0 & 0.891  \\
                            ~ & MEEG-ResNet3D-17 &\textbf{93.0} & \textbf{93.0} & \textbf{93.2} & 0.894  \\
                            ~ & MEEG-ResNet3D-21 & 91.8 & 91.8 & 92.4 & \textbf{0.900}  \\\hline
       
	\end{tabular} 
}
	\label{tab:hyperparams}
\end{table} 


 \begin{table}[tb]
    \small
	\centering 
	\caption{Performance of MEEG-ResNet3D with different sensor permutation strategies for intra- and inter-cluster processing.}
  \resizebox{9.0cm}{1.2cm}{
	\begin{tabular}{lccccc}
		\hline 
		\textbf{Datasets} & \textbf{Permutation}
		&\textbf{TPR ($\%$)} &\textbf{TNR ($\%$)}&\textbf{F1 ($\%$)}&\textbf{AUROC}   \\ \hline 
		\multirow{3}*{Sanbo-CMR} & left & 93.92 & 93.09 & 93.91 & 0.984  \\
                            ~ & right & 94.52 & 93.17 & 94.51 & 0.984  \\
                            ~ & center & \textbf{94.73} & \textbf{94.73} & \textbf{94.73} & \textbf{0.988} \\ 
                            ~ & inter-cluster & 94.35$\pm$ 0.103 & 94.11$\pm$ 1.274 & 94.35$\pm$ 0.096 & 0.982$\pm$ 0.0024 \\ \hline
        \multirow{3}*{TUSZ}  & left & \textbf{93.0} & \textbf{93.0} & \textbf{93.2} & \textbf{0.894}  \\
                            ~ & right & 92.8 & 92.8 & 93.0 & 0.856 \\ 
                            ~ & inter-cluster & 90.8$\pm$ 1.166 & 90.7$\pm$ 1.187 & 91.7$\pm$ 0.770 & 0.888$\pm$ 0.006  \\\hline
	\end{tabular} } 
	\label{permutation results}
\end{table}
\subsection{Qualitative Analysis on the Spike Detection Task}
\begin{figure}[tb]
    \centering
    \includegraphics[width=9cm,height=2.5cm]{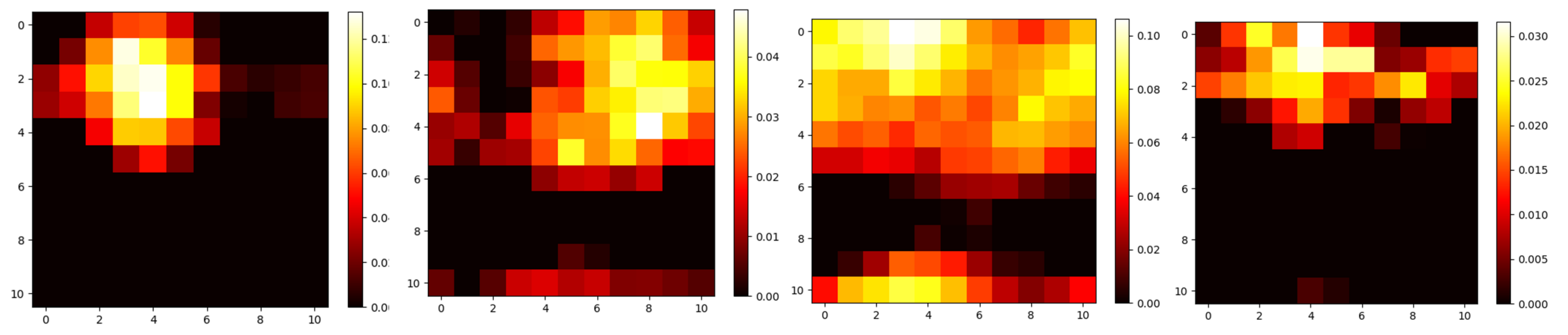}
    \caption{Four activation maps from the test set of \textit{Sanbo-CMR}, each corresponding to a spatial-cluster map in the 3D input. }
    \label{fig:result_analysis1}
\end{figure}

\begin{figure}[tb]
   \centering
  \includegraphics[scale=.25]{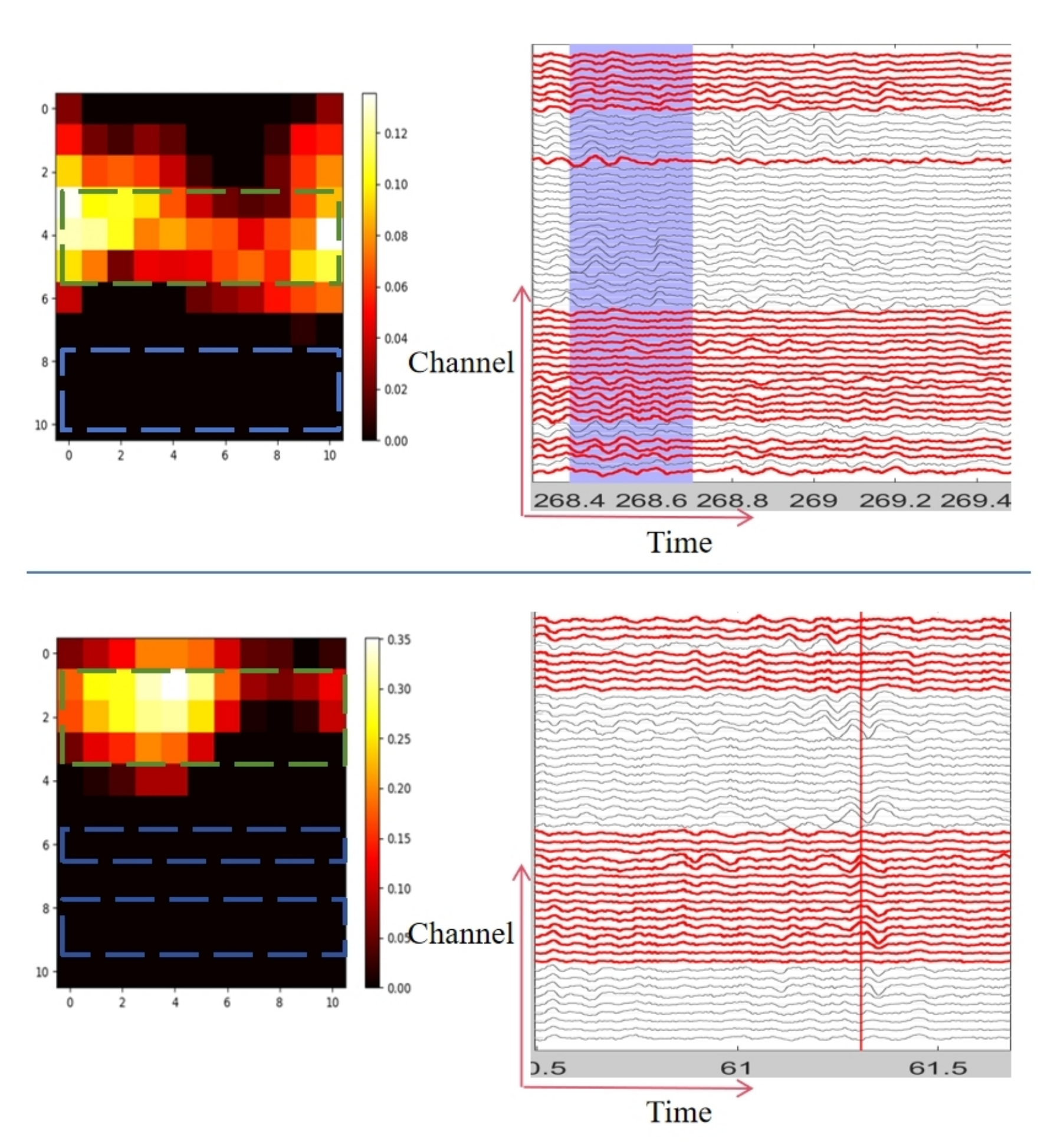}
    \caption{Two exemplary samples from the test set of \textit{Sanbo-CMR}. The areas inside the green rectangle align with the red channels from three clusters, while the regions within the blue rectangle correspond to the gray channels.}
    \label{fig:result_analysis2}
\end{figure}

Utilizing Grad-CAM \cite{selvaraju2017grad}, we generate activation maps of the spatial-cluster map within the input to identify the regions that significantly affect the prediction result. Fig. \ref{fig:result_analysis1} presents exemplary activation maps where black regions indicate that the corresponding channel signals do not contribute to the label prediction whatsoever. It is thereby evident that only certain channel signals within specific clusters determine the label category, aligning with regional spiking character of the signals introduced at the beginning of this paper. To confirm that this phenomenon was not accidental, we manually assessed the activation maps of 71 test samples and found that 76\% of them exhibited a situation where only partially connected regions contributed to the prediction result.  Additionally, these activation regions can be projected back into time-series, 
aiding in the identification of signal characteristics, as illustrated in Fig. \ref{fig:result_analysis2}. The signals corresponding to the activated regions in the activation map indeed exhibit spiking characteristics at a certain time point or within a time range. Both figures demonstrate that our model is capable of capturing regional spiking characteristics and thereby enhancing spike detection performance. The input design introduced in Section \ref{input encoding} also enhances our model's interpretability as we can determine the spiking locations when projecting the activated regions into their Euclidean coordinates mapped for the brain.

\section{Conclusions and Future Work}
\label{sec:future work}

In this study, we introduce a cutting-edge MEEG-ResNet3D model, enhanced with a novel spatial clustering prior, for effective seizure analysis. This model exhibits superior performance compared to multiple state-of-the-art approaches in both epileptic MEG spike detection and EEG seizure detection. Extensive experiments are conducted to identify the factors that contribute to the model's outstanding performance, particularly the fusion of spatial-clustering priors with the MEEG-ResNet3D architecture. Our method boasts high interpretability, as it aids in identifying spike locations on the head. In future research, we plan to explore strategies for improving coordinate establishment to enhance clustering and investigate the integration of additional data, such as frequency domain inputs and scalp maps, to further improve our model's performance.

\bibliographystyle{ieeetr}
\bibliography{main}

\end{document}